\documentclass[universe,article,oneauthor,10pt,a4paper]{mdpi} 
\usepackage{subcaption}
\usepackage{amsmath,amsfonts,amssymb}
\preto{\abstractkeywords}{\nolinenumbers}
\firstpage{1} 

\Title{Prospects of constraining the dense matter equation of state from the timing analysis of pulsars in double neutron star binaries: the cases of PSR J0737$-$3039A and PSR J1757$-$1854}

\Author{Manjari Bagchi }

\address{The Institute of Mathematical Sciences (IMSc-HBNI), 
4th Cross Road, CIT Campus, Chennai, India 600 113; manjari.bagchi@gmail.com\\
}

\abstract{The Lense-Thirring effect from spinning neutron stars in double neutron star binaries contribute to the periastron advance of the orbit. This extra term involves the moment of inertia of the neutron stars. Moment of inertia, on the other hand, depends on the mass and spin of the neutron star as well as the equation of state of the matter. If at least one member of the double neutron star binary (better the faster one) is a radio pulsar, then accurate timing analysis might lead to the estimation of the contribution of the Lense-Thirring effect to the periastron advance, which will lead to the measurement of the moment of inertia of the pulsar. Combination of the knowledge on the values of the moment of inertia, the mass, and the spin of the pulsar, will give a new constraint on the equation of state. Pulsars in double neutron star binaries are the best for this purpose as short orbits and moderately high eccentricities make the Lense-Thirring effect substantial, whereas tidal effects are negligible (unlike pulsars with main sequence or white-dwarf binaries). The most promising pulsars are PSR J0737$-$3039A and PSR J1757$-$1854. The spin-precession of pulsars due to the misalignment between the spin and the orbital angular momentum vectors affect the contribution of the Lense-Thirring effect to the periastron advance. This effect has been explored for both  PSR J0737$-$3039A and PSR J1757$-$1854, and as the misalignment angles for both of these pulsars are small, the variation in the Lense-Thirring term is not much. However, to extract the Lense-Thirring effect from the observed rate of the periastron advance, more accurate timing solutions including precise proper motion and distance measurements are essential.}


\begin{document}
%%%%%%%%%%%%%%%%%%%%%%%%%%%%%%%%%%%%%%%%%%

%%%%%%%%%%

\section{Introduction}

Timing analysis of binary pulsars lead to the measurement of pulsar's spin, Keplerian orbital parameters (the orbital period $P_{\rm b}$, the orbital eccentricity $e$, the longitude of periastron $\omega$, the projected semi-major axis of the orbit $x_p = a_p \, sin \, i$ where $a_p$ is the semi-major axis of the pulsar orbit and $i$ is the angle between the orbit and the sky-plane, and the epoch of the periastron passage\footnote{The remaining Keplerian parameter, the longitude of the ascending node $\varphi$ does not come in the standard pulsar timing algorithm, it can be measured via proper motion only in very special cases, see \cite{vbb01, dbl13, dvk16} for details.}), as well as post-Keplerian (PK) parameters like the Einstein parameter ($\gamma$), Shapiro range ($r$) and shape ($s$) parameters, the rate of the periastron advance $\dot{\omega}$, the rate of change of the orbital period $\dot{P}_{\rm b}$, the relativistic deformation of the orbit $\delta_{\theta}$, etc. \citep{hbopa}. Sometimes, the Shapiro delay is parametrized differently, with parameters $h_3$ and $\varsigma$ \cite{fw10}, or with $r$ and $z_s$ \cite{ksm06a}, and these parameters can easily be expressed in terms of conventional parameters $s$ and $r$.

Measurement of PK parameters leads estimation of masses of the pulsars and their companions, as well as tests of various theories of gravity \citep{st03, kr16}. Note that, in principle, measurements of only two PK parameters are enough to extract two unknowns, i.e., the masses of the pulsar and the companion while measurements of more than two PK parameters lead tests of gravity theory through consistency. However, the uncertainty in measurement is not equal for every PK parameter and usually the two most accurate parameters are used to obtain the best mass estimates (e.g., Fig 1 of \cite{ksm06b}). Additionally, relativistic binary pulsars have the potential to constrain the Equation of State (EoS) of matter at extreme densities. Measurements of masses of two pulsars being around $2~M_{\odot}$ have already ruled out many soft EsoS \cite{dpr10, afw13}. On the other hand, the recent analysis of the gravitational wave event GW170817 from the merger of two neutron stars ruled out some extremely stiff EsoS \citep{ligoDNS17, agk17}. Still a large number of EsoS are allowed, including several hybrid \cite{abb17} and strange quark matter \cite{krv10} EsoS in addition to standard hadronic ones. So, further progress in this issue is essential, and that is expected in the near future as in one hand more neutron star$-$neutron star mergers are expected to be detected in future runs of the advanced LIGO, and on the other hand, many binary pulsars are being timed regularly and accurately. Besides, upcoming radio telescopes like MeerKAT and SKA will lead significant improvement in pulsar timing.

Among all of the PK parameters, in the present article, I concentrate mainly on the periastron advance, which has a potential to constrain the dense matter equation of state due to the Lense-Thirring effect. Neutron star$-$neutron star binaries or `double neutron star' (DNS) systems are the best for this purpose as these systems are compact enough to display effects of strong field gravity yet wide enough to have negligible tidal effects. Moreover, due to high compactness of neutron stars, spin induced quadrupole moments are also negligible. On the other hand, as these neutron stars are rapidly spinning (spin periods of most of the pulsars in DNSs are less than 100 milliseconds), the Lense-Thirring effect is significant. That is why I concentrate on DNSs in this article, although the mathematical formulations are valid for any kind of general relativistic binaries with a possibility of additional effects depending on the nature of the objects in the binaries. There are about sixteen DNSs known at the present time, including the first discovered binary pulsar - the Hulse-Taylor binary, PSR B1913$+$16. For one DNS, PSR J0737$-$3039A/B, both members are radio pulsars and the system is known as the double pulsar \footnote{Although for the last few years, the slow pulsar of the system is not visible and is believed to be beaming away from the earth due to its spin-precession \citep{pmk10}.}. 

For a long period of time, the double pulsar was the most relativistic binary. However, the recently discovered DNS, PSR J1757$-$1854 shows even stronger effects of general relativity in some aspects, i.e., $\dot{P}$ and $\gamma$ \citep{cck17} due to its high eccentricity combined with a small orbital period. In fact, its eccentricity is 6.9 times and orbital period is 1.8 times larger than those of PSR J0737$-$3039A/B. Noteworthily, $\dot{\omega}$ is larger for PSR J0737$-$3039A/B than PSR J1757$-$1854 where the contribution of eccentricity is less dominant, see \citet{hbopa} for expressions for PK parameters.

PSR J1906$+$0746 is another DNS having the second smallest value of $P_b$, just after PSR J0737$-$3039A/B. However, its $P_b$ is slightly (1.1 times) and $e$ is significantly (7.1 times) smaller than those of PSR J1757$-$1854. That is why although PSR J1906$+$0746 has the third largest value of $\dot{\omega}$, other PK parameters are small, even smaller than most of the other DNSs. Finally, the latest discovered DNS, PSR J1946$+$2052, has broken all records by having the smallest value of $P_b$ and the largest value of $\dot{\omega}$ \citep{sfc18}.

\section{Precession in double neutron star binaries} 

Precessions (of both the spin and the orbit) of neutron stars in DNSs are very important. Following \citet{bo79}, the rate of the change of the unit spin vector (${\bf s}_a$) of a spinning neutron star ($a$) in a binary can be written as:

\begin{equation}
{\bf \dot{s}}_a = \overrightarrow{\Omega}_{{\rm s}a} \times {\bf s}_a ~,
\label{eq:spinchange}
\end{equation} where the angular spin-precession frequency $\overrightarrow{\Omega}_{{\rm s}a}$ can be written as:

\begin{equation}
\overrightarrow{\Omega}_{{\rm s} a} = \overrightarrow{\Omega}_{{\rm sPN}_a} + \overrightarrow{\Omega}_{{\rm sLT}_a}   ~.
\label{eq:spinrecor}
\end{equation} 

\noindent $ \overrightarrow{\Omega}_{{\rm sPN}_a} $ is the angular spin-precession frequency of the neutron star due to the space-time curvature around its companion and can be calculated within the `post-Newtonian' (PN) formalism. $\overrightarrow{\Omega}_{{\rm sLT}_a} $ is the angular spin-precession frequency of that neutron star due to the Lense-Thirring effect of its spinning companion. We can further write:

\begin{equation}
 \overrightarrow{\Omega}_{{\rm sPN}_a} = A_{{\rm PN}_a} \, {\bf k}  ~,
 \label{eq:PNspin}
\end{equation} where ${\bf k} = \overrightarrow{L}/ \vert \overrightarrow{L} \vert$ is the unit vector along the orbital angular momentum, $\overrightarrow{L}$ is the orbital angular momentum, and

\begin{equation}
\overrightarrow{\Omega}_{{\rm sLT}_a} = A_{{\rm LT}_a} \, \left[ {\bf s} _{a+1}- 3 ( {\bf k}  . {\bf s}_{a+1}) {\bf k}  \right]  ~.
\label{eq:LTspin}
\end{equation} 

Here if $a$ is the neutron star under consideration, $a+1$ is its companion. The amplitudes in Eqns. \ref{eq:PNspin} and \ref{eq:LTspin} are given by:

\begin{equation}
A_{{\rm PN}_a} = \left( \frac{G}{c^3} \right)^{2/3} \, \frac{n^{5/3}}{(1-e^2)} \, \frac{M_{a+1} (4 M_a + 3 M_{a+1})}{2 (M_a + M_{a+1})^{4/3}} ~,
\label{eq:PNspinampl}
\end{equation}

\begin{equation}
 A_{{\rm LT}_a} = \frac{G}{c^3} \, \beta_{s \, a+1} \, \frac{n^2}{(1-e^2)^{3/2}} \, \frac{M_{a+1}^2}{2 (M_a + M_{a+1} ) } ~,
 \label{eq:LTspinampl}
\end{equation} where $G$ is the gravitational constant, $c$ is the speed of light in vacuum, and $n~=~2\pi/P_{\rm b}$ is the angular orbital frequency. $M_a$ is the mass, $I_a$ is the moment of inertia, $P_{s a}$ is the spin period of the $a$-th neutron star, and

\begin{equation}
\beta_{s \, a+1 }~=~\frac{c I_{a+1} }{G M_{a+1}^2} \cdot \frac{2 \pi}{P_{s \, a+1}} ~, ~~~ \beta_{sa}~=~\frac{c I_a }{G M_a^2} \cdot \frac{2 \pi}{P_{sa}} ~.
\label{eq:betas}
\end{equation}

For pulsars in DNSs, the Lense-Thirring term ($\overrightarrow{\Omega}_{{\rm sLT}_a}$) can be ignored. Even for the case of the slow pulsar (B) of the double pulsar, where the Lense-Thirring effect due to the spin of the fast pulsar (A) contributes to $\overrightarrow{\Omega}_{\rm s_B}$, I find $A_{\rm PN_B} = 5.07481 ~{\rm deg~yr^{-1}}$ and $A_{\rm LT_B} = 3.48865 \times 10^{-5}~{\rm deg~yr^{-1}}$ (using the values of the parameters given in Table \ref{tab:DNSsreala}). This leads to ${\Omega}_{\rm sLT_B} \sim - 6.97731  \times 10^{-5}  ~{\rm deg~yr^{-1}}$ using the fact that ${\bf s}_{\rm A}$ is almost parallel to ${\bf k}$ \citep{pkm14, pmk18}, and hence ${\Omega}_{\rm s_B} = 5.07474 ~{\rm deg~yr^{-1}}$, which is close to the observed median value of ${\Omega}_{\rm s_B} = 4.77 ~{\rm deg~yr^{-1}}$ \citep{bkk08}. It is unlikely that companions of other pulsars would be much faster than PSR A, and hence $\overrightarrow{\Omega}_{{\rm sLT}_a}$ can be neglected. So, Eqn. \ref{eq:spinchange} becomes

\begin{equation}
{\bf \dot{s}}_a = A_{\rm PN_a} \, {\bf k} \times {\bf s}_a  =  A_{\rm PN_a} \, \sin \chi_a ~ {\bf u} ~,
\label{eq:spinchange2}
\end{equation} where $\chi_a$ is the angle between ${\bf k}$ and ${\bf s}_a$ and ${\bf u}$ is a unit vector perpendicular to the plane containing ${\bf k}$ and ${\bf s}_a$ in the direction given by the right-hand rule. So, the measurement of the spin-precession of a pulsar helps estimate $\chi_a$ when other parameters involved in Eqn. \ref{eq:PNspinampl} are already known.

Similarly, neglecting the tidal and the spin-quadrupole effects, as well as the spin-spin interaction, the orbital angular precession frequency can be written as:
 
\begin{equation}
 \overrightarrow{\Omega}_{{\rm b}} = \overrightarrow{\Omega}_{{\rm bPN}} + \overrightarrow{\Omega}_{{\rm bLT}} ~,
\label{eq:orbin:precessfreq}
\end{equation} where $  \overrightarrow{\Omega}_{{\rm bPN}}$ and $\vec{\Omega}_{{\rm b LT}}$ are contributions from the space-time curvature and the Lense-Thirring effect respectively. Note that, both members of the binary contribute to each term. $\overrightarrow{\Omega}_{{\rm b}}$ leads to the precession of both the Laplace-Runge-Lenz vector $\vec{\mathcal{A}}$ as well as $\vec{L}$, but none is directly observable. \citet{ds88} first studied the manifestation of $ \overrightarrow{\Omega}_{{\rm b}}$ in terms of observable parameters by decomposing $ \overrightarrow{\Omega}_{{\rm b}}$ as:

\begin{equation}
\vec{ \Omega}_{{\rm b}} = \frac{d \varphi_a}{dt} {\bf h}_a + \frac{d \omega_a}{dt} {\bf k} + \frac{d i}{dt} {\bf \Upsilon}_a ~,
\end{equation} where ${\bf h}_a$ is the unit vector along the line-of-sight, i.e., from the earth to the $a$-th neutron star (the pulsar), ${\bf \Upsilon}_a = \frac{{\bf h}_a \times {\bf k}}{\vert {\bf h}_a \times {\bf k}\vert}$ is the unit vector along the line of the ascending node, ${\varphi}_a $ is the longitude of the ascending node of the $a$-th neutron star and $i$ is the angle between the orbit and the sky-plane. \citet{ds88} has also given:

\begin{subequations}
\begin{equation}
\dot{\varphi}_a  = \frac{1} {\sin^2 i} [\vec{ \Omega}_{{\rm b}} . {\bf h}_a - \cos i (\vec{ \Omega}_{{\rm b}} . {\bf k})] ~,
\label{eq:ds88eq2A}
\end{equation}
\begin{equation}
\dot{\omega}_a =  \frac{1} {\sin^2 i} [\vec{ \Omega}_{{\rm b}} . {\bf k} - \cos i (\vec{ \Omega}_{{\rm b}} . {\bf h}_a)] ~,
\label{eq:ds88eq2B}
\end{equation}
\begin{equation}
\frac{d i}{d t} = \vec{ \Omega}_{{\rm b}} . {\bf  \Upsilon}_a ~.
\label{eq:ds88eq2C}
\end{equation}
\end{subequations} As already mentioned, $\dot{\omega}_a$ is the parameter of interest as it has the potential to put constraints on the dense matter EoS. So, I explore properties of $\dot{\omega}_a$ in the next and subsequent sections.

\subsection{Periastron advance}

\begin{figure}
    \centering
    \begin{subfigure}[b]{0.48\textwidth}
        \includegraphics[width=\textwidth]{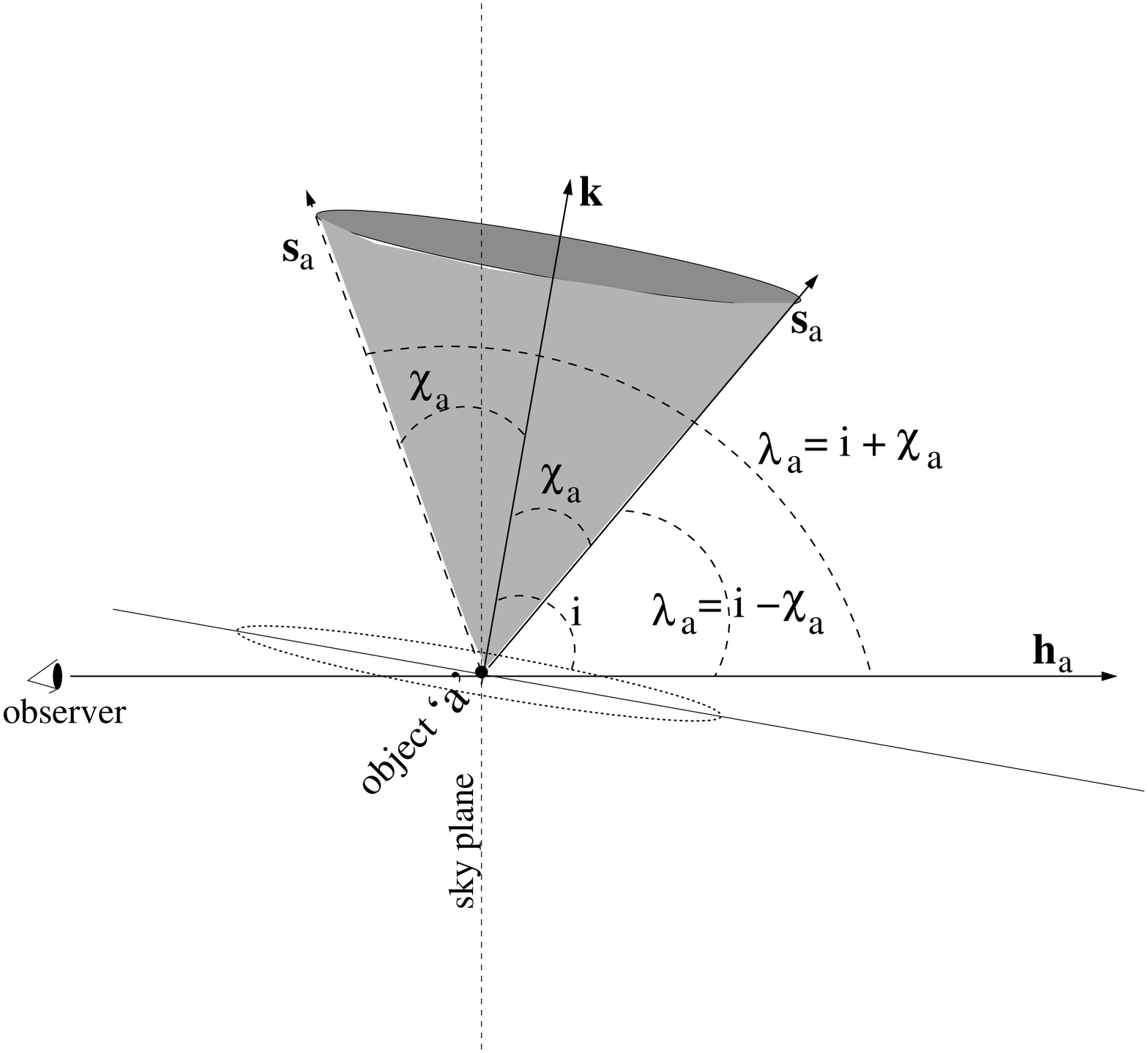}
        \caption{$\chi_a < i$}
        \label{fig:geometryA}
    \end{subfigure}
    \quad    
    \begin{subfigure}[b]{0.48\textwidth}
        \includegraphics[width=\textwidth]{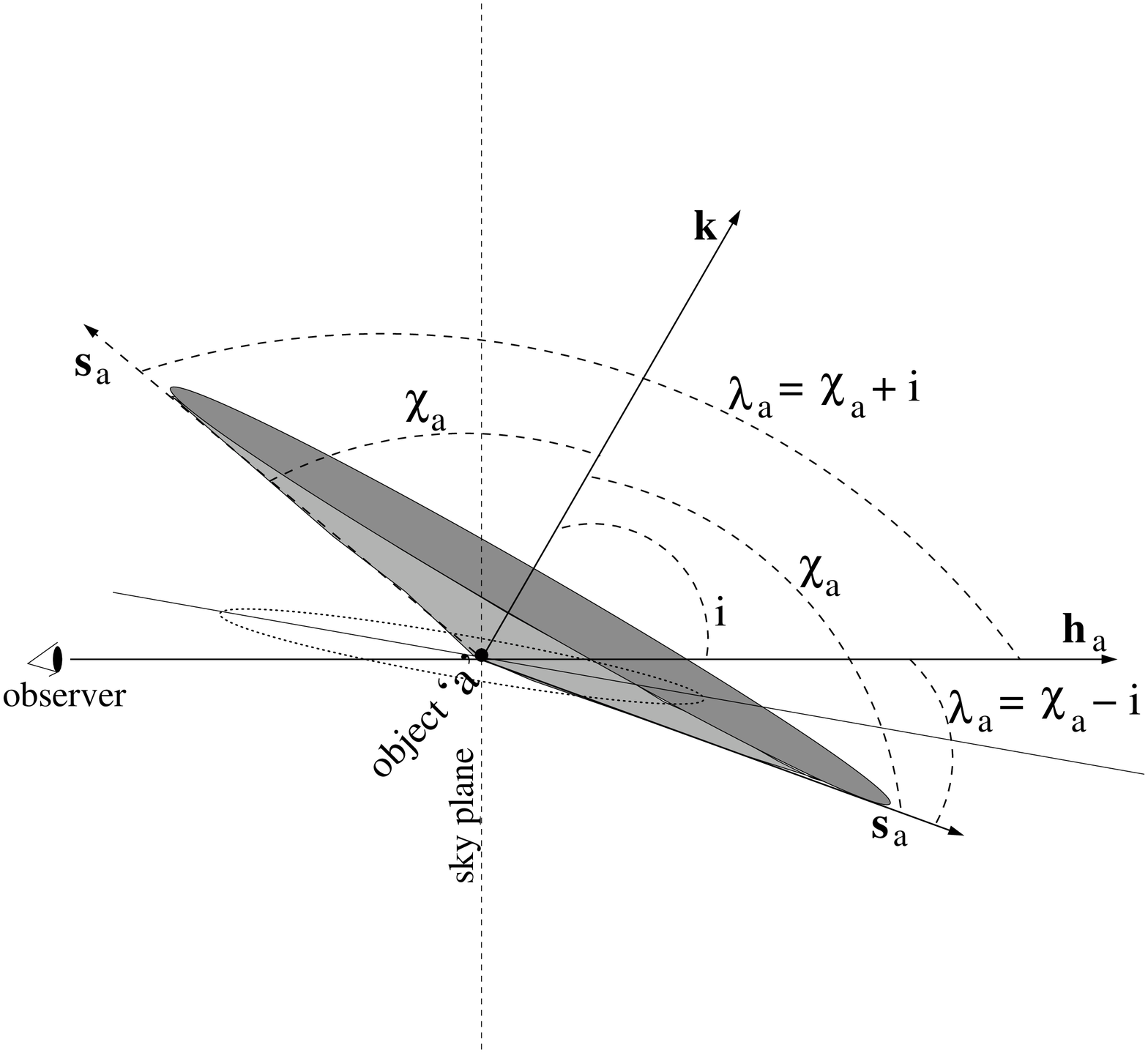}
        \caption{$\chi_a > i$}
        \label{fig:geometryB}
    \end{subfigure}
        \caption{Orientation of different vectors relevant for the estimation of $\dot{\omega}_{{\rm LT} a}$ where the subscript $a$ refers to the object being observed, i.e., the pulsar (so the companion would be represented by the subscript $a+1$). The vectors are as follow: ${\bf k}$ is the unit vector along the orbital angular momentum, ${\bf h}_a$ is the unit vector along the line-of-sight, and ${\bf s}_a$ is the unit spin vector. The angles in the figures are: $i$ is the inclination angle between the orbit of the pulsar and the sky-plane as well as the angle between ${ \bf   k}$ and ${ \bf   h}_a$, $\chi_a$ is the angle between ${\bf k}$ and ${\bf s}_a$, and $\lambda_a$ is the angle between ${ \bf   h}_a$ and ${\bf s}_a$. The left and right panels are for $\chi_a < i$ and $\chi_a > i$, respectively. In both of the panels, two positions of ${\bf s}_a$ are shown corresponding to maximum and minimum values of $\lambda_a $ (see text). The pulsar, i.e., the object $a$ is located at the vertex of the fudicial cone (gray in color) made by the precessing ${ \bf   s}_a$ around ${\bf k}$.}\label{fig:geometry}
\end{figure}

The observed rate of the periastron advance ($\dot{\omega}_{a, {\rm obs}}$) of the $a$-th member of a relativistic binary can be written as:

\begin{equation}
\dot{\omega}_{a, {\rm obs}}~ = \dot{\omega}_{a} +  \dot{\omega}_{{\rm Kop}_a} = \dot{\omega}_{{\rm PN}_a} + \dot{\omega}_{{\rm LT}_a} + \dot{\omega}_{{\rm LT}_{a+1}} + \dot{\omega}_{{\rm Kop}_a}
\label{eq:per_adv}
\end{equation} where $\dot{\omega}_{{\rm PN}_a}$ is due to the space-time curvature caused by both members of the binary, $\dot{\omega}_{{\rm LT}_a}$ is due to the Lense-Thirring effect of the star $a$, $\dot{\omega}_{{\rm LT}_{a+1}}$ is due to the Lense-Thirring effect of the star $a+1$. These three terms together comes into the expressions of $\dot{\omega}_{a}$ (Eqn. \ref{eq:ds88eq2B}) caused by the precession of the orbit. $\dot{\omega}_{{\rm Kop}_a}$ is a secular variation due to the gradual change of the apparent orientation of the orbit with respect to the line-of-sight due to the proper motion of the barycenter of the binary \citep{kop96}. The expressions for different terms are:
\begin{subequations}
\begin{equation}
\dot{\omega}_{{\rm PN}_a} =  ~\frac{3 \beta_0^2~ n}{1-e^2} [1 + \beta_0^2 f_{0a}] ~,
\label{eq:per_adv2a}
\end{equation}
\begin{equation}
\dot{\omega}_{{\rm LT}_a} + \dot{\omega}_{{\rm LT}_{a+1}} = -  ~ \frac{3 \beta_0^3 ~ n }{1-e^2} (g_{sa} \beta_{sa}+g_{s \, a+1} \beta_{s \, a+1}) ~,
\label{eq:per_adv2b}
\end{equation}
\begin{equation}
\dot{\omega}_{{\rm Kop}_a} = \mathcal{K}  ~ {\rm cosec} \, i \, \left( \mu_{\alpha} \cos {\varphi}_{a} +  \mu_{\delta} \sin {\varphi}_{a} \right) ~,~ \mathcal{K} = 502.65661 ~,
\label{eq:per_adv2c}
\end{equation}
\label{eq:per_adv2}
\end{subequations} where only the first and the second order terms are retained in the expression of $\dot{\omega}_{{\rm PN}_a}$. $\mu_{\alpha}$ and $ \mu_{\delta}$ are the proper motion of the barycenter of the binary (which is measured as the proper motion of the visible object) in the right ascension and the declination respectively, both expressed in the units of milliarcseconds per year. All other parameters are in the SI units. The parameters introduced in Eqns. \ref{eq:per_adv2}a,b,c are as follow:

\begin{equation}
\beta_0~=~\frac{(GMn)^{1/3}}{c}  ~,
\label{eq:beta0}
\end{equation}
\begin{equation}
f_{0a}~=~\frac{1}{1-e^2}\left( \frac{39}{4}X_a^2+\frac{27}{4}X_{a+1}^2+15 X_a X_{a+1} \right)  - \left( \frac{13}{4} X_a^2+\frac{1}{4} X_{a+1}^2+\frac{13}{3} X_a X_{a+1} \right) ~,
\label{eq:f0}
\end{equation} where $X_a~=~M_a/M$, $M = M_a + M_{a+1}$ is the total mass of the system. $\beta_{sa}$ is defined in Eqn. \ref{eq:betas}, and

\begin{eqnarray}
g_{sa}~=\frac{X_a \left(4 X_a+ 3X_{a+1}\right)}{6(1-e^2)^{1/2} {\rm sin^2} i} \times \left[ (3 ~{\rm sin^2} i-1)~{ \bf   k}~.~{\bf s}_a+{\rm cos} \, i~ { \bf h}_a~.~{ \bf s}_a \right] \nonumber \\
=\frac{X_a \left(4 X_a+ 3X_{a+1}\right)}{6(1-e^2)^{1/2} {\rm sin^2} i} \times \left[ (3 ~{\rm sin^2} i-1)~ {\rm cos} \, \chi_a +{\rm cos} \, i~ {\rm cos} \, \lambda_a \right] 
\label{eq:gs}
\end{eqnarray} where $\lambda_a$ is the angle between ${ \bf   h}_a$ and ${\bf s_a}$ (see Fig. \ref{fig:geometry}). It is obvious from Eqn. \ref{eq:gs} that the maximum value of $g_{sa}$ occurs when ${\bf s}_a$ is parallel to the vector $(3~ {\rm sin^2 } i-1)~{\bf   k}+{\rm cos}~ i~ { \bf h}_a $, giving
\begin{equation}
 {g_{sa, ~max}}=~\left[3+\frac{1}{{\rm sin ^2} ~i} \right]^{1/2}\frac{X_a \left(4 X_a+ 3 X_{a+1}\right)}{6(1-e^2)^{1/2}} ~.
\label{eq:gsamax}
\end{equation}
If ${\bf s}_a~\Vert~ {\bf k}$, i.e., $\chi_a=0$ and $\lambda_a = i$:
\begin{equation}
g_{sa,~\Vert}=~\frac{X_a \left(4 X_a+ 3 X_{a+1}\right)}{3(1-e^2)^{1/2}} ~.
\label{eq:gs_par}
\end{equation}

Using the expression of $g_{sa,~\Vert}$ in Eqn. \ref{eq:per_adv2b}, I get $ \dot{\omega}_{{{\rm LT}_a}, \Vert} $, and using the expression of $g_{sa,~max}$, I get $ \dot{\omega}_{{\rm LT}_a, ~max} $. The negative sign in the Lense-Thirring term (Eqn. \ref{eq:per_adv2b}) implies the fact that actually $\dot{\omega}_{{\rm LT}_a, ~max} $ is the minimum value. However, depending on the values of $i$, $\chi_p$, $\lambda_p$; $g_{sa}$ (Eqn \ref{eq:gs}) can be negative making $\dot{\omega}_{{\rm LT}_a}$ (Eqn \ref{eq:per_adv2b}) positive. Note that, the Lense-Thirring effects from both the pulsar and the companion come in the total periastron advance rate (Eqns. \ref{eq:per_adv}, \ref{eq:per_adv2b}). But if the companion is much slower than the pulsar, as in the case of the double pulsar, then the contribution from the companion can be neglected. Due to the non-detection of any pulsation, we do not know the values of the spin periods of the companions for other DNSs, and cannot rule out the possibility of significant Lense-Thirring effect from those companions. However, most of the pulsars in DNSs are recycled (at least mildly), suggesting the fact that pulsars were born as neutron stars earlier than their companions. The companions, i.e., the second born neutron stars in DNSs, are expected to be slow; because even if neutron stars are born with spin periods in the range of a few tens of milliseconds up to a few hundred milliseconds, they quickly spin down to periods of a few seconds unless they gain angular momentum via mass accretion from their Roche Lobe filling giant companions and become fast rotators. Such spin-up is possible only for the first born neutron star in a DNS. That is why in the present article, I ignore the Lense-Thirring effect from the companions of pulsars in DNSs. 

In such a case, i.e. when $\dot{\omega}_{{\rm LT}_{a+1}}$ is negligible, $\dot{\omega}_{{\rm LT}_a}$ can be extracted by subtracting $\dot{\omega}_{{\rm PN}_a} + \dot{\omega}_{{\rm Kop}_a}$ from $\dot{\omega}_{a, {\rm obs}}$ (Eqn. \ref{eq:per_adv}). One can estimate the value of $I_a$ from $\dot{\omega}_{{\rm LT}_a}$ if all other relevant parameters like $P_b$, $e$, $M_{a}$, $M_{a+1}$, $\sin i$, $\chi_a$, $\lambda_a$ are known (Eqns \ref{eq:betas}, \ref{eq:beta0}, \ref{eq:f0}, and \ref{eq:gs}). Then, from the knowledge of $I_a$, $M_a$, and $P_{s a}$, a new constraint on the EoS can be placed as it is a well-known fact that the moment of inertia of a neutron star depends on its mass, spin-period and the EoS \citep[and references therein]{bag10}. For this purpose, DNSs with large values of $\dot{\omega}_{{\rm LT}_a}$ (larger than the measurement uncertainty in $\dot{\omega}_{a, {\rm obs}}$) are preferable. Large values of the orbital eccentricity $e$, and small values of $P_{\rm b}$ and $P_{s}$ make $\dot{\omega}_{{\rm LT}_a}$ larger. 

The procedure is actually a bit trickier than it seems first. Usually, timing analysis of a binary pulsar reports the most accurate measurement of $\dot{\omega}_{a, {\rm obs}}$ out of all PK parameters, and the masses of the pulsar and the companion are estimated by equating $\dot{\omega}_{a, {\rm obs}}$ with $\dot{\omega}_{{\rm PN}_a}$ and using the second most accurately measured  PK parameter. If $\dot{\omega}_{{\rm LT}_a}$ is large, then this procedure would become erroneous. One should instead estimate masses using two PK parameters other than $\dot{\omega}_{a, {\rm obs}}$, and these mass values should also be accurate enough to give a precise estimate of $\dot{\omega}_{{\rm PN}_a}$. Accurate measurements of proper motion is also needed to evaluate $\dot{\omega}_{{\rm Kop}_a}$. Only then, $\dot{\omega}_{{\rm PN}_a} + \dot{\omega}_{{\rm Kop}_a}$ can be subtracted from the observed $\dot{\omega}_{a, {\rm obs}}$ to get the value of  $\dot{\omega}_{{\rm LT}_a}$. In short, one needs very precise measurements of at least three PK parameters, $\dot{\omega}_{a, {\rm obs}}$ and two more, as well as the distance and the proper motion of the pulsar (the $a$-th neutron star).

Because of its small values of $P_{\rm b}$ and $P_s$, PSR J0737$-$3039A is expected to have a large value of $\dot{\omega}_{{\rm LT}_a}$ in spite of its low eccentricity. Similarly, as PSR J1757$-$1854 has a slightly larger $P_{\rm b}$, slightly smaller $P_s$, and much larger $e$, it is also expected have a large value of $\dot{\omega}_{{\rm LT}_a}$. For this reason, in the next section, I explore $\dot{\omega}_{{\rm LT}_a}$ for these two DNSs in detail. Note that, although, PSR J1946$+$2052 might be a useful system for this purpose, it is not possible presently to extend my study for this system due to the lack of knowledge of masses of the pulsar and the companion.

One should also remember the dependence of $ \dot{\omega}_{{\rm LT}_a}$ (via $g_{sa}$) on the orientation of ${\bf s}_a$ with respect to ${\bf k}$ and ${\bf h}_a$ (Eqn \ref{eq:gs}). If I assume that the $a$-th neutron star is the pulsar, then all `$a$'s in the subscripts of above equations will be replaced by `$p$'s, e.g., ${\bf s_p}$, ${\bf h_p}$, etc. Most of the time, the value of $g_{sp,~\Vert}$ is estimated in the literature. However, at least for some pulsars $\chi_p \ne 0$. In such cases, the values of $\chi_p$ and $\lambda_p$ can be measured by analyzing the change of the pulse profile shape and the polarization. Conventionally, $\chi_p$ is denoted by $\delta$ and $\lambda_p$ is denoted by $\zeta$. Note that, even if $\chi_p$ is a constant over time, $\lambda_p$ would vary due to the spin precession and can be written as \citep{fsk13}: 

\begin{equation}
 \cos \lambda_p = - \cos \chi_p \, \cos i  + \sin  \chi_p \, \sin i \, \cos[\Omega_{sp} (t -T_{p0})] ~.
\label{eq:lamdavariation}
\end{equation} where $\Omega_{sp}$ is the angular spin precession frequency (amplitude of $\overrightarrow{\Omega}_{{\rm s} p}$ in Eqn. \ref{eq:spinrecor}), $t$ is the epoch of the observation, $T_{p0}$ is the epoch of the precession phase being zero. Note that, $i$ itself might change with time as given in Eqn. (\ref{eq:ds88eq2C}).

In Table \ref{tab:DNSsreala}, with other relevant parameters, I compile values of $\chi_p$ wherever available. I do not compile values of $\lambda_p$ although available in some cases as it is a time-dependent parameter. From Fig. \ref{fig:geometry}, it is clear that the spin would precess in such a way that it would lie on the surface of a fiducial cone with the vertex at the pulsar and having the half-opening angle as $\chi_p$. The constraint on $\lambda_p$ is: (i) if $\chi_p < i$, then $\lambda_{p, max} = i + \chi_p$ and $\lambda_{p, min} =  i - \chi_p$ (Fig. \ref{fig:geometryA}), (ii) if $\chi_p > i$, then $\lambda_{p, max} = \chi_p + i$ and $\lambda_{p, min} =   \chi_p - i$ (Fig. \ref{fig:geometryB}).

%%%%%%%%%%%%%%%%%%%%%%%%%%%%%%%%%%%%%%%%%%
\section{Results}

For the purpose of demonstration, I compute the value of the moment of inertia ($I_p$) for the two most interesting pulsars, PSR J0737$-$3039A and PSR J1757$-$1854 using the APR equation of state \citep{apr98} and the RNS code\footnote{RNS stands for `Rapidly Rotating Neutron Star', a package to calculate different properties of rotating neutron stars, freely available at http://www.gravity.phys.uwm.edu/rns/.}\citep{nsge98}. I find that, both of these pulsars have $I_p = 1.26 \times 10^{45} ~{\rm gm~ cm^2}$. I use this value in my calculations, remembering the fact that the true values of $I_p$ would be different depending on the true EoS and we are actually seeking an answer whether it would be possible to know the value of $I_p$ by singling out $\dot{\omega}_{{\rm LT}_p}$ from the total observed $\dot{\omega}_{p, {\rm obs}}$.

I tabulate values of $\dot{\omega}_{{{\rm PN}_p}}$, $ \dot{\omega}_{{{\rm LT}_p}, \Vert} $, and $ \dot{\omega}_{{\rm LT}_p, ~max} $ for all pulsars in DNSs in Table \ref{tab:DNSsrealb}. I exclude candidate DNSs, i.e., PSR B1820$-$11 and PSR J1753$-$2240 from this calculation due to poor mass constraints. For other cases, I use the limiting mass values where actual values are unavailable. Only $   \dot{\omega}_{{{\rm LT}_p}, \Vert}  $ can be calculated for the systems with undetermined/unpublished values of $\sin i$. As expected, PSR J0737$-$3039A shows the largest Lense-Thirring effect, followed by PSR J1757$-$1854. In fact, $\vert \dot{\omega}_{{\rm LT}_p, ~max} \vert/ \vert \dot{\omega}_{{{\rm PN}_p}} \vert$ is slightly larger for PSR J1757$-$1854 ($2.92 \times 10^{-5}$) than that for PSR J0737$-$3039A ($2.81 \times 10^{-5}$). This fact makes it a very interesting system, and I investigate this system in detail. The discovery paper \citep{cck17} also mentioned large Lense-Thirring effect causing significant amount of $di/dt$ (Eqn. \ref{eq:ds88eq2C}).

\begin{figure}
 \centering
           \includegraphics[width=0.6\textwidth, angle=-90]{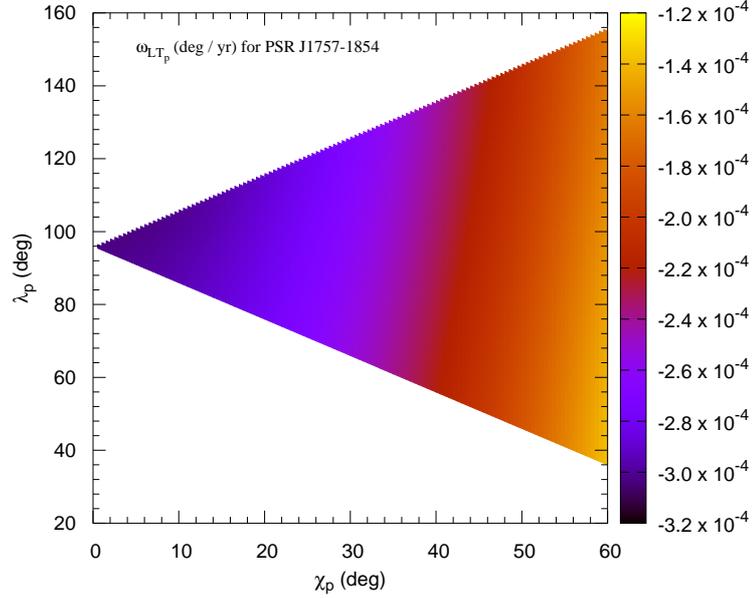}          
        \caption{Variation of $\dot{\omega}_{{\rm LT}_p}$ with $\chi_p$ (in degrees) along the horizontal axis and $\lambda_p$ (in degrees) along the vertical axis for PSR J1757$-$1854. The color code represents values of $\dot{\omega}_{{\rm LT}_p}$ in ${\rm deg~ yr^{-1}}$.}\label{fig:omLTJ0757}
\end{figure}

I find that, for PSR J1757$-$1854, $\vert \dot{\omega}_{{\rm LT}_p, ~max}  \vert = 3.03170 \times 10^{-4}~{\rm deg~yr^{-1}}$ is achieved for $\chi_p = 3^{\circ}$, $\lambda_p = 99^{\circ}$. But as the value of $\chi_p$ for this system is not yet known (although \citet{cck17} obtained a constraint $\chi_p \sim 25^{\circ}$ based on their simulation of the kick velocity), I vary $\chi_p$ in the range of $0^{\circ} - 60^{\circ}$. It is obvious that $\dot{\omega}_{{\rm LT}_p} $ depends more on $\chi_p$ than on $\lambda_p$ as the absolute value of the factor (see Eqn. \ref{eq:gs}) with $\cos \chi_p$, i.e., $3 \sin^2 i-1$ is larger than that of the factor with $\cos \lambda_p$, i.e., $ \cos i $. Also, the measured value of $\sin i$ gives two values of $i$ (as $\sin \theta = \sin (\pi - \theta) $), I use the larger one to get a bigger range of $\lambda_p$. The minimum value of $\vert \dot{\omega}_{{\rm LT}_p} \vert = 1.37597 \times 10^{-4}~{\rm deg~yr^{-1}}$ is achieved for $\chi_p = 60^{\circ}$, $\lambda_p = 36^{\circ}$. If I fix $\chi_p$ strictly as $25^{\circ}$, then the maximum and minimum values of $\vert \dot{\omega}_{{\rm LT}_p} \vert$ are $2.81108 \times 10^{-4}~{\rm deg~yr^{-1}}$ and $2.67663  \times 10^{-4}~{\rm deg~yr^{-1}}$, for $\lambda_p = 121^{\circ}$ and $\lambda_p = 71^{\circ}$, respectively. In Fig. \ref{fig:omLTJ0757}, I show the variation of $\dot{\omega}_{{\rm LT}_p}$ for $\chi_p$ in the range of $0^{\circ} - 60^{\circ}$, and for each value of $\chi_p$, $\lambda_p$ varies in the range of $i - \chi_p, ~  i + \chi_p$. On the other hand, as PSR J0737$-$3039A has $\chi_p < 6^{\circ}$ \citep{{pkm14}}, $\dot{\omega}_{{\rm LT}_p}$ always remains close to $  \dot{\omega}_{{{\rm LT}_p},  \parallel} $. The maximum and minimum values of $\vert \dot{\omega}_{{\rm LT}_p} \vert$ are $4.74488 \times 10^{-4}~{\rm deg~yr^{-1}}$ at $\chi_p=0.65$, $\lambda_p=91.95^{\circ}$; and $4.712931 \times 10^{-4}~{\rm deg~yr^{-1}}$ at $\chi_p=6.0^{\circ}$, $\lambda_p=85.31^{\circ}$. To check the validity of my assumption that the Lense-Thirring effect from the companion could be ignored, I find that $\dot{\omega}_{{\rm LT}_c, ~ \parallel}  = -3.792899 \times 10^{-6}~{\rm deg~yr^{-1}}$ using $I_c = 1.16 \times 10^{45} ~{\rm gm~ cm^2}$ where the subscript `$c$' stands for the companion, i.e., the slow PSR J0737$-$3039B. Varying $\chi_c$ in the range of $130^{\circ} - 150^{\circ}$ \cite{pkm14}, I find that the minimum value of $\vert \dot{\omega}_{{\rm LT}_c} \vert$ is $  2.60052  \times 10^{-6}~{\rm deg~yr^{-1}}$. These results support my argument that the Lense-Thirring effect from the companion of a pulsar in a DNS would have negligible contribution to the rate of the periastron advance of the binary.

As Table \ref{tab:DNSsrealb} shows, $\vert \dot{\omega}_{{\rm LT}_p, ~max} \vert$ for PSR J0737$-$3039A is around 1.4 times smaller than the presently published accuracy, $\dot{\omega}_{p, {\rm obs}} = 16.89947 \pm 0.00068~{\rm deg~{yr^{-1}}}$ where the uncertainty is twice the formal $1\sigma$ value obtained in the timing solution \citep{ksm06b}. The uncertainties in other PK parameters are even poorer (Table 1 of \citep{ksm06b}), so if we exclude $\dot{\omega}_{p, {\rm obs}}$, the mass estimates would be less accurate and should not be used to calculate $\dot{\omega}_{{\rm PN}_p}$. So, to achieve the goal of estimating $\dot{\omega}_{{\rm LT}_p}$, lowering only the uncertainty in $\dot{\omega}_{p, {\rm obs}}$ would not be enough, one needs to improve the accuracy of at least two other parameters that can be used to get masses at least as precise as the ones already published. It is impossible to improve the accuracy of the Keplerian parameter $R$ used by \citet{ksm06b} in combination with $\dot{\omega}_{p, {\rm obs}}$ to report the values of the masses, because $R=M_{p}/M_{c} = x_{c}/x_{p}$ involves $x_c$, the projected semi-major axis of the companion (PSR B), which is not visible presently and even when it was visible, it was not a good timer as PSR A. Solving the expressions for PK parameters, I find that the combination of $s$ and $\dot{P}_b$ is the best choice left when $\dot{\omega}_{p, {\rm obs}}$ and $R$ are excluded, i.e., gives the narrowest ranges of masses $M_p = 1.34 \pm 0.02 ~M_{\odot}$, $M_c= 1.251 \pm 0.007~M_{\odot}$ using the published values of the uncertainties. I also find the fact that, to get masses as accurate as the ones reported by \citet{ksm06b}, the uncertainty in $\dot{P}_b$ should be reduced at least to $ 8.0 \times 10^{-16}$ assuming that $s$ and relevant Keplerian parameters ($P_b$, $e$, $x_p$) would also be improved by at least an order of magnitude. The published uncertainty in $\dot{P}_b$ is $1.7 \times 10^{-14}$. However, the question arises whether for such an improved measurement of $\dot{P}_b$, the dynamical contribution coming from the proper motion of the pulsar (Shklovskii effect) and the relative acceleration between the pulsar and the solar system barycenter, could be ignored. Using the proper motion reported by \citet{ksm06b} and the distance (1.1 kpc) estimated by \citet{dbt09} using VLBI parallax measurement, I find that $\dot{P}_{b,~ {\rm dyn}} = 4.91 \times 10^{-17}$ where I have used a realistic model of the Galactic potential \cite{pb17}. The improvement required in the timing solution of PSR J0737$-$3039A is expected by 2030 \citep{kw09, kwkl16} accompanied by an improvement in $\dot{\omega}_{p, {\rm obs}}$. Fortunately, $\dot{\omega}_{{\rm Kop}_p}$ is smaller than $\vert \dot{\omega}_{{\rm LT}_p, ~ \parallel} \vert$ - using the proper motion reported in \citet{ksm06b}, I find that it can be at most $1.18 \times 10^{-6} ~{\rm deg~{yr^{-1}}}$ (I could not calculate the actual value, as $\varphi_p$ is unknown). However, one will still need better estimates of $\chi_p$ and $\lambda_p$ to be able to extract the value of the moment of inertia of the pulsar from the $\dot{\omega}_{{\rm LT}_p}$.

On the other hand, for PSR J1757$-$1854, both $\vert \dot{\omega}_{{\rm LT}_p, ~max} \vert$ and $\vert \dot{\omega}_{{\rm LT}_p, ~ \parallel} \vert$ are about 1.5 times larger than the presently published accuracy, $\dot{\omega}_{p, {\rm obs}} = 10.3651 \pm 0.0002~{\rm deg~{yr^{-1}}}$. However, even this system cannot be used at present to estimate $\dot{\omega}_{{\rm LT}_p}$, not only because of the unknown values of $\chi_p$ and $\lambda_p$, but also because of the less accurate values of other PK parameters. If one uses $\gamma$ and $\dot{P}_b$, then the uncertainties in these parameters should be improved by one and two orders of magnitudes respectively to get the uncertainty in masses of about $0.0008 ~ M_{\odot}$. This seems plausible, as the present accuracy in $\dot{P}_b$ is only $0.2 \times 10^{-12}$ \cite{cck17}. However, due to the lack of measurements of the proper motion and the parallax distance, I could not calculate $\dot{P}_{b, ~{\rm dyn}}$ for this pulsar, only can give rough estimates based on the distance guessed from the dispersion measure using two different models of the Galactic electron density, e.g., NE2001 model \cite{cl02, cl03} and YMW16 model \cite{ymw17}. The contribution due to the relative acceleration is $-2.11 \times 10^{-15}$ for the NE2001 distance (7.4 kpc), and $-1.65 \times 10^{-14}$ for the YMW16 distance (19.6 kpc). Both of the values seem to be too large to be canceled out by the Shlovskii term. Measurements of the proper motion and the distance (using parallax) are essential to estimate $\dot{P}_{b, ~{\rm dyn}}$. The measurement of the proper motion will also help estimate $\dot{\omega}_{{\rm Kop}_p}$.

\begin{table*}\footnotesize
\begin{center}
 \setlength\tabcolsep{3.5pt}
\caption{Relevant observed parameters for known pulsars in DNSs. See references for these parameters with more significant digits and uncertainties. Timing model (e.g., DDGR) is mentioned when more than one timing solutions are available in the original reference.}
\begin{tabular}{|l|l|c|c|c|c|c|c|c|c|}
\hline\hline
 DNS & Masses & $e$ & $P_{\rm b}$  & $\sin i$ & $P_{s1}$   & $\dot{\omega}$ & $\chi_p$ & $~$ Refs.$~$ \\
  &  &  &  &  &   &   & & \\
  & ${\rm M_{\odot}}$ &  & days  &  & ms  &  ${\rm ~ deg~ yr^{-1}}$ & deg &\\
\hline

 J0453$+$1559 & 1.559 &  0.11251832 & 4.072468588  &  0.9671343 & 45.781816163093   &  0.0379412 & $-$ & R1\\
 (DDGR) & 1.174 &  &  & &  &  & &\\
 &  &  &  &  &   &  & &\\

 J0737$-$3039A & $1.3381$ & 0.0877775 & 0.10225156248 & 0.99974 &  22.70   & 16.89947 & $< 6.1 ~{\rm or}~< 2.3$ & R2, R$\chi_1$ \\
J0737$-$3039B & $1.2489$ &  &  & & 2773.46  & & 138 &\\
 &  &  &  &  &  &   & &\\

 J1411$+$2551 & $\leq 1.62$ & 0.1699308 &  2.61585677939 & $-$ &  62.452895517590  & 0.0768 & $-$ & R3\\
  &  $\geq 0.92$ &  &  & &  &  & &\\
 &  &  &  &  &  &   & &\\

 J1518$+$4904   & $\leq 1.17$ &  0.24948451 &  8.6340050964 & $ \leq 0.73$  & 40.934988908  &  0.0113725 & $-$ & R4\\
 & $\geq 1.55$ &  &  & &  &  & & \\
 &  &  &  &  &  &   & &\\

 B1534$+$12  & $1.3455$ &   0.27367740 &  0.420737298881 & 0.97496   & 37.9044411783  &  1.755795 & $27.0 \pm 3.0$ & R5 \\
 (DDGR) & $1.3330$ &  &  &   &    & & $153.0 \pm 3.0$ &  \\
 &  &  &  &  &   &  & & \\

 J1753$-$2240$^\dagger$  & $-$ &   0.303582 &  13.6375668 & $-$ &  95.1378086771   & $-$ & $-$ & R6\\
 & $\geq 0.4875$ &  &  &  &  &   & &\\
 &  &  &  &  &   & & &\\

 J1756$-$2251  &   1.341 &  0.1805694 &   0.31963390143 & 0.93  &   28.4615890259983 &    2.58240  & $< 5.9$ & R7\\
 &  1.230 &  &  &  &  &  & ($1\sigma$ value) &\\
 &  &  &  &  &  &   & &\\
 
  J1757$-$1854 & 1.3384  &  0.6058142 &  0.18353783587  & 0.9945 &  21.497231890027   & 10.3651 & $\sim 25$ & R8  \\
 & 1.3946 &  &  &  & & & (theoretical) & \\
 &  &  &  &  &  &  & &\\

 J1807$-$2500B$^\dagger$  & 1.3655 &  0.747033198 &  9.9566681588  &  0.9956  & 4.18617720284089   &  0.0183389  & $-$ & R9 \\
 $(g_1)$ &  1.2064 &  &  &  &  & &  &\\
  &  &  &  &  &  &  & &\\
 
 J1811$-$1736  & $\leq 1.64 $ & 0.828011 & 18.7791691  & $-$ &   104.1819547968 &  0.0090 & $-$ & R10 \\
  & $\geq 0.93$ &  &  &  & &  &  &\\
  &  &  &  &  &  &  & &\\
  
 B1820$-$11$^\dagger$  & $-$ &  0.794608 & 357.76199 & $-$ &  279.828696565349 &  0.00007 & $-$ & R11\\
  &  &  &  &  &  &   & &\\
  &  &  &  &  &  &   & &\\

 J1829$+$2456 & $\leq 1.38$ & 0.13914 & 1.176028 & $-$ &  41.00982358 &  0.28 & $-$ & R12\\
    & $\geq 1.22$ &  &  &  &   &  & &\\
   &  &  &  &  & &   & &\\

 J1906$+$0746 & 1.291 & 0.0852996  &  0.16599304686 &  0.690882411 & 144.07315538    &   7.5844 & $110^{+21}_{-55}$ & R13  \\
(DDGR)  &  1.322 &  &  &  &  &   & &\\
 &  &  &  &  & &  & &\\

 B1913$+$16 & $  1.4398 $ &  0.6171334 &   0.322997448911 & 0.68$^{\ddagger}$ &  59.03000322  &   4.226598 & 22 & R14, R$\chi_1$ \\
 & $1.3886$ &  &  &  &  & & &\\
 &  &  &  &  &   &  & &\\

  J1930$-$1852 & $\leq 1.32$  &  0.39886340  & 45.0600007  & $-$ &   185.52016047926 & 0.00078  & $-$ & R15  \\
  & $\geq 1.30$  &  &  & &  & &  & \\
  &  &  &  &  &  &  & &\\
  
   J1946$+$2052 & $-$  &  0.063848  & 0.07848804  & $-$ &  16.9601753230  & 25.6  & $-$ & R16  \\
  &   &  &  & &  & &  & \\
  &  &  &  &  &  &  & &\\

 B2127$+$11C & 1.358  &  0.681395 &   0.33528204828 & $-$ &  30.52929614864 &  4.4644 & $-$ & R17 \\
 $(g_2)$ & 1.354 &  &  &  &  & &  & \\
  &  &  &  &  &  &   & &\\

\hline
\end{tabular}
{\footnotesize $^\dagger$ candidate DNS. \\ }
{\footnotesize $^{\ddagger}$ was never published, but here I quote the value used by \citet{wh16}. \\ }
{\footnotesize pulsars denoted with $g$ are in globular clusters, $g_1$ is in NGC 6544, and $g_2$ is in M15. \\ }
{\footnotesize{References : \\ R1: \citet{msf15}, R2: \citet{ksm06b}, R3: \citet{msf17}  , R4: \citet{jsk08},  R5: \citet{fst14}, \\ R6:  \citet{kkl09}, R7: \citet{fsk14}, R8:  \citet{cck17}, R9: \citet{lfrj12}, R10 \citet{cks07}, \\ R11: \citet{hlk04}, R12: \citet{clm04}, R13: \citet{vks15}, R14:  \citet{wnt10}, R15: \citet{srm15}, R16: \citet{sfc18}, R17: \citet{jcj06}, R$\chi_1$ (for $\chi_p$): \citet{pkm14}.  }}
\label{tab:DNSsreala}
\end{center}
\end{table*}

\begin{table*}
\begin{center}
\caption{Values of $   \dot{\omega}_{{\rm PN}_p } $, $\dot{\omega}_{{\rm LT}_p, ~\parallel} $ and $ \dot{\omega}_{{\rm LT}_p, ~max} $ for known pulsars in DNSs using $I_{p} = 1.26 \times 10^{45} ~{\rm gm~ cm^2}$ and neglecting the Lense-Thirring effect of the companion. To show the difference between $ \dot{\omega}_{{{\rm LT}_p}, \Vert} $ and $ \dot{\omega}_{{\rm LT}_p, ~max} $, I keep values up to five decimal places even for the cases where $\sin i$ is known with less accuracy.}
\begin{tabular}{lccc}
\hline\hline
DNS    & $   \dot{\omega}_{{\rm PN}_p } $   & $  \dot{\omega}_{{\rm LT}_p, ~\parallel} $ & $  \dot{\omega}_{{\rm LT}_p, ~max} $    \\
    & ${\rm (deg~ yr^{-1})}$ & ${\rm (deg~ yr^{-1})}$   & ${\rm (deg~ yr^{-1})}$    \\
\hline
J0453$+$1559  & 0.03794 &  $-1.30190  \times 10^{-7} $ &   $-1.31310  \times 10^{-7 } $   \\ \\

J0737$-$3039A & 16.90312 &  $-4.74458 \times 10^{-4}$ &  $-4.74489 \times 10^{-4} $    \\  \\

J1411$+$2551   & 0.07681 &  $-2.32510 \times 10^{-7} $ &   $-$   \\  \\

J1518$+$4904   & 0.01138 &  $-4.48005 \times 10^{-8} $ &  $- $ \\  \\

B1534$+$12     & 1.75533 &  $-1.84880 \times 10^{-5} $ &  $-1.86078 \times 10^{-5} $  \\  \\

J1753$-$2240   & $- $ &  $- $ &   $- $   \\  \\

J1756$-$2251   & 2.58363 &  $-4.01930 \times 10^{-5} $ &  $-4.09703 \times 10^{-5} $ \\  \\

J1757$-$1854   & 10.36772 & $-3.02752 \times 10^{-4} $ & $-3.03170 \times 10^{-4}$   \\  \\

J1807$-$2500B   & 0.01834 & $-8.98082 \times 10^{-7}$ & $-8.99076 \times 10^{-7}$     \\  \\

J1811$-$1736    & 0.00895 &  $-1.45038 \times 10^{-8}$ &  $-$  \\  \\

B1820$-$11     & $- $ &  $ - $  & $- $ \\  \\

J1829$+$2456   & 0.29284 & $-1.96704 \times 10^{-6}$  & $- $ \\  \\

J1906$+$0746   & 7.58528 &  $-2.91884 \times 10^{-5}$  &  $-3.29423 \times 10^{-5} $ \\  \\

B1913$+$16    & 4.22760 &  $-3.43984 \times 10^{-5} $  & $-3.90790 \times 10^{-5} $  \\  \\

J1930$-$1852 & 0.00079 &  $-3.86975 \times 10^{-10}$ &   $- $   \\  \\

J1946$+$2052   & $- $ &  $ - $  & $- $ \\  \\

B2127$+$11C  & 4.46458 &  $-8.11241 \times 10^{-5 }$  & $- $  \\  \\
\hline
\end{tabular}
\label{tab:DNSsrealb}
\end{center}
\end{table*}

%%%%%%%%%%%%%%%%%%%%%%%%%%%%%%%%%%%%%%%%%%
\section{Summary and Conclusions}

Measurement of the moment of inertia of a pulsar will provide yet another constraint on the dense matter equation of state. This issue has been discussed in the literature in the past \cite[and references therein]{iorio09, oc04, kwkl16}, but unfortunately the effect of the misalignment between the spin and the orbital angular momentum vectors has not been explored in detail mainly because these two vectors are almost parallel for PSR J0737$-$3039, the only suitable system existed until recently.

However, even a small misalignment angle would have significant consequences, because for a particular neutron star, the theoretical values of the moment of inertia using different EsoS are sometimes very close (depending on the stiffness of the EoS). One can see Fig 1 of \citet{bbh05} for predicted values of the moment of inertia for PSR J0737$-$3039A using a sample of EsoS. The present article extensively studied the effects of such misalignment.

I have demonstrated the fact that the recently discovered DNS, PSR J1757$-$1854 is almost as good as PSR J0737$-$3039A regarding the prospect of determining the moment of inertia of the pulsar from $\dot{\omega}_{{\rm LT}_p} $. For PSR J0737$-$3039A, $\dot{\omega}_{{\rm LT}_p} $ lies in the range of $ - 4.757 \times 10^{-4}$ to $ - 4.713 \times 10^{-4}$ ${\rm deg~ yr^{-1}}$ whereas for PSR J1757$-$1854 $ \dot{\omega}_{{\rm LT}_p} $ lies in the range of $-1.376 \times 10^{-4}$ to $-3.032 \times 10^{-4}$ ${\rm deg~ yr^{-1}}$. Future measurement of $\chi_p$ for PSR J1757$-$1854 will narrow down the range of $\dot{\omega}_{{\rm LT}_p}$.

For both of the systems, significant improvements in timing solution are needed. Moreover, the proper motion and the parallax measurements for PSR J1757$-$1854 is crucial. Additionally, it is  essential to measure $\chi_p$ and $\lambda_p$ for PSR J1757$-$1854, hence long-term studies of polarization properties and profile variation would be very useful.

Note that, although all of the calculations presented in this article is within the framework devised by \citet{bo79} and \citet{ds88}, alternative formalisms to incorporate the Lense-Thirring effect in $\dot{\omega}_{p, {\rm obs}}$ exist. One example is \citet{io12}, but for systems with $i \sim 90^{\circ}$, $\chi_p \sim 0^{\circ}$ these two formalisms take similar forms. Precisely, \citet{io12} reports $ \dot{\omega}_{{\rm LT}_p, ~ \parallel} = -3.7 \times 10^{-4}~{\rm deg \, yr^{-1}}$ for PSR J0737$-$3039A using $I_p = 1.00 \times 10^{45} ~{\rm gm~ cm^2}$, which is close to what I get, i.e. $ \dot{\omega}_{{\rm LT}_p, ~ \parallel} = -3.766 \times 10^{-4}~{\rm deg \, yr^{-1}}$ if I use the same value of $I_p$ instead of $1.26 \times 10^{45} ~{\rm gm~ cm^2}$ used in this article. A fundamentally different idea, i.e., to incorporate the Lense-Thirring effect into the delays in the pulse arrival times has been proposed recently \cite{io17}. If this can be implemented in future algorithms for pulsar timing, the effect of the Lense-Thirring effect could be measured directly.

Finally, if the measurement accuracy of $\dot{\omega}_{p, {\rm obs}}$ improves by a few orders of magnitudes, one will need to subtract the Kopeikin term (the secular variation due to the proper motion). The measurement of the proper motion of PSR J1757$-$1854 will help us calculate this term. Moreover, in order to use $\dot{P}_b$ to estimate masses accurately, one needs to subtract the contributions from the proper motion and the relative acceleration between the pulsar and the solar system barycenter. A package to perform these tasks based on a realistic potential of the Milky Way has been recently developed \citep{pb17}, and gradual improvement in the model of the Galactic potential is expected with the help of Gaia data.

%%%%%%%%%%%%%%%%%%%%%%%%%%%%%%%%%%%%%%%%%%
%% optional

%%%%%%%%%%%%%%%%%%%%%%%%%%%%%%%%%%%%%%%%%%
\section{Acknowledgement}

The author thanks the organizers of CSQCD VI (especially Prof. David Blashcke) for an excellent hospitality, all three reviewers for extensive suggestions on the earlier version of the manuscript, and ANI Technologies Pvt. Ltd. and Uber Technologies Inc. for helping survive in Chennai.


%=====================================
% References, variant A: internal bibliography
%=====================================
\reftitle{References}

%%%%%%%%%%%%%%%%%%%%%%%%%%%%%%%%%%%%%%%%%%
\end{document}